\documentclass{article} 
\usepackage{iclr2026_conference,times}

\usepackage{amsmath,amsfonts,bm}

\def\eqref#1{equation~\ref{#1}}

\def\1{\bm{1}}

\DeclareMathAlphabet{\mathsfit}{\encodingdefault}{\sfdefault}{m}{sl}
\SetMathAlphabet{\mathsfit}{bold}{\encodingdefault}{\sfdefault}{bx}{n}

\usepackage{hyperref}
\usepackage{url}

\usepackage{amsmath}
\usepackage{mathtools}
\usepackage{siunitx}
\usepackage{booktabs}
\usepackage{graphicx}
\usepackage{wrapfig}
\usepackage{subcaption}

\title{Boltzmann Generators for Condensed Matter via Riemannian Flow Matching}

\author{Emil Hoffmann, Maximilian Schebek, Leon Klein\\
Freie Universität Berlin\\
Berlin, Germany \\
\texttt{\{emil.hoffmann,m.schebek,leon.klein\}@fu-berlin.de} \\
\And
Frank Noé \\
Microsoft Research \& Freie Universität Berlin\\
Berlin, Germany \\
\texttt{frank.noe@microsoft.com} \\
\And
Jutta Rogal \\
Flatiron Institute \\
New York, USA \\
\texttt{jrogal@flatironinstitute.org} \\
}
\iclrfinalcopy 
\begin{document}

\maketitle
\lhead{Published as a workshop paper at AI4MAT, ICLR 2026}
\begin{abstract}
Sampling equilibrium distributions is fundamental to statistical mechanics. While flow matching has emerged as scalable state-of-the-art paradigm for generative modeling, its potential for equilibrium sampling in condensed-phase systems remains largely unexplored.
We address this by incorporating the periodicity inherent to these systems into continuous normalizing flows using Riemannian flow matching.
The high computational cost of exact density estimation intrinsic to continuous normalizing flows is mitigated by using Hutchinson's trace estimator, utilizing a crucial bias-correction step based on cumulant expansion to render the stochastic estimates suitable for rigorous thermodynamic reweighting.
Our approach is validated on monatomic ice, demonstrating the ability to train on systems of unprecedented size  and obtain highly accurate free energy estimates without the need for traditional multistage estimators.

\end{abstract}
\section{Introduction}
Accurate estimation of thermodynamic observables, such as free energy and heat capacity, in condensed matter systems relies fundamentally on the efficient sampling of the high-dimensional equilibrium distributions of large systems.
Traditionally, molecular dynamics (MD) and Monte Carlo (MC) simulations are considered as primary sampling tools, but their sequential nature requires many steps to generate uncorrelated samples, making them computationally expensive. In recent years, machine learning approaches have emerged as a promising alternative. In particular, Boltzmann Generators (BGs)~\citep{noe_boltzmann_2019} leverage invertible neural networks, such as normalizing flows, to generate equilibrium samples directly while enabling tractable likelihood evaluation.

Continuous normalizing flows (CNFs)~\citep{chen_neural_2018} trained with the scalable and efficient flow-matching training paradigm~\citep{lipman_flow_2023, albergo2023building, liu2023flow} have emerged as the state-of-the-art approach for modeling complex biomolecular systems~\citep{klein_transferable_2024, rehman_falcon_2025}. However, the development of flexible CNFs for condensed-phase systems, commonly modeled using periodic boundary conditions, remains limited~\citep{grenioux_riemannian_2025}. Previous studies have primarily employed architecturally constrained coupling-flow models~\citep{dinh_density_2017} for crystals~\citep{wirnsberger_normalizing_2022,schebek_efficient_2024,ahmad_free_2022} and liquids~\citep{wirnsberger_targeted_2020,coretti_learning_2025}, but these approaches were limited by computational cost, restricting training to system sizes  of a few hundred particles, which is significantly smaller than what is required to eliminate finite-size effects~\citep{polson_finitesize_2000}.

The goal of this work is to introduce CNFs with periodic boundary conditions in combination with Riemannian flow matching~\citep{chen_flow_2024} and tractable density estimation as a flexible, scalable, and size-transferable framework for equilibrium sampling of condensed-phase systems. We show that this approach overcomes the scalability limitations of prior coupling-flow architectures by training on a monatomic ice system more than four times larger than previous benchmarks, achieving excellent sample quality while matching the computational budget of the smaller baselines.

\section{Methods}

\textbf{Boltzmann Generators} combine exact-likelihood deep generative models, such as normalizing flows~\citep{rezende_variational_2015, papamakarios_normalizing_2021}, with importance reweighting to sample from a target Boltzmann distribution $\mu(\mathbf{x}) = Z^{-1} \exp[-\beta U(\mathbf{x})]$.
This distribution is defined by the potential $U(\mathbf{x})$, the inverse temperature $\beta=(k_BT)^{-1}$, with Boltzmann constant $k_B$, temperature $T$ and the partition function $Z$.
The model is trained to learn an invertible mapping $f_\theta$ that transforms a prior sample $\mathbf{x}_0 \sim p(\mathbf{x}_0)$ into a generated sample $\mathbf{x}_1 = f_\theta(\mathbf{x}_0)$, whose distribution $\tilde{p}(\mathbf{x}_1)$ approximates $\mu(\mathbf{x}_1)$.
Unbiased estimates can be obtained by reweighting the generated samples to the target distribution using the importance weights 
\begin{equation}
    w(\mathbf{x}_0)  =  \exp\left(\beta_p U_p(\mathbf{x}_0) - \beta_\mu U_\mu(f_\theta(\mathbf{x}_0)) - \Delta \log p \right),
\label{eq:weights}
\end{equation}
such that $w(\mathbf{x}_0)\propto \mu(f_\theta(\mathbf{x}_0))/\tilde{p}(f_\theta(\mathbf{x}_0))$. The density change $\Delta \log p$ is defined as  
\begin{equation}
\Delta \log p := \log \tilde{p}(f_{\theta}(\mathbf{x}_0)) - \log p(\mathbf{x}_0) = - \log \left| \det J_{f}(\mathbf{x}_0) \right|, 
\end{equation}
where $J_f$ denotes the Jacobian of the transformation.
The importance weights further provide a measure of modeling performance via the effective sample size (ESS)~\citep{kish_sampling_1965} defined by ${\rm ESS}=(\sum_{i=1}^{N_S} w(\mathbf{x}_0^{(i)}))^2/\sum_{i=1}^{N_S} (w(\mathbf{x}_0^{(i)}))^2$, which is typically reported relative to the number of samples $N_S$.
While the free energy $F$, which governs phase stability, is defined in terms of the partition function $Z$ as $F = -k_B T \log Z$, evaluating the partition function is typically intractable.
Within the framework of  targeted free energy perturbation (TFEP)~\citep{jarzynski_targeted_2002}, a trained BG can be used to directly estimate the free energy difference between prior and target states through $\beta\Delta F_{p \mu} = -\log\mathbb{E}_{\mathbf{x}_0 \sim p(\mathbf{x}_0)} \left[ w(\mathbf{x}_0) \right]$. This allows the calculation of free-energy differences without sampling intermediate states via MD simulations, which are required by traditional estimators that rely on sufficient phase-space overlap between distributions~\citep{shirts_statistically_2008}. Knowing the free energy of the prior state allows estimation of the absolute free energy of the target system and, consequently, a comparison of the thermodynamic stability of different phases.

\textbf{Normalizing Flows} learn an invertible mapping $f_\theta$ to transform samples from a prior distribution $p(\mathbf{x}_0)$ to a generated density $\tilde{p}(\mathbf{x}_1)$ that approximates the target distribution $\mu(\mathbf{x}_1)$.
 CNFs~\citep{chen_neural_2018} model this transformation in continuous time by solving an ordinary differential equation (ODE) over $t \in [0,1]$,
\begin{equation}
\frac{d f_\theta^t(\mathbf{x})}{dt} = v_\theta(t, f_\theta^t(\mathbf{x})), \quad f_\theta^0(\mathbf{x}) = \mathbf{x}_0,
\end{equation}
where $v_\theta$ is a time-dependent vector field with learnable parameters $\theta$. The resulting density evolution follows the continuous change-of-variables formula 
\begin{equation}
    \Delta \log p = - \int_{0}^{1} \nabla \cdot v_\theta(t, f_\theta^t(\mathbf{x}_0)) \, dt,
\end{equation}
where the  divergence of the vector field is equivalent to the trace of the vector field's Jacobian.
CNFs can be naturally extended to general Riemannian manifolds $\mathcal{M}$ by formulating and solving the associated ODE directly on $\mathcal{M}$~\citep{mathieu_riemannian_2020}. 

\textbf{Riemannian Flow Matching.} 
While normalizing flows are commonly trained by maximum-likelihood objective, 
\citet{lipman_flow_2023} introduced flow matching for efficient and simulation-free training of CNFs, which has been generalized to manifolds by~\citet{chen_flow_2024}.
In the case of periodic boundary conditions, corresponding to the flat torus, the Riemannian conditional flow matching (RFM) objective allows to train the vector field through
\begin{align}
    \mathcal{L}_{\mathrm{RFM}}(\theta) = \mathbb{E}_{t \sim [0,1],\, \mathbf{x} \sim p_t(\mathbf{x}|\mathbf{z})} \left[ \| v_\theta(t, \mathbf{x}) - u_t(\mathbf{x}|\mathbf{z}) \|_2^2 \right],
\end{align}
where  $p_t(\mathbf{x}|\mathbf{z})$ is a probability path conditioned on an arbitrary distribution $\mathbf{z}$. A proven choice for this path and the conditional distribution $\mathbf{z}$ is the direct constant speed interpolation between a single prior and target sample~\citep{chen_flow_2024}, which yields the conditional vector field
\begin{align}
    u_t(\mathbf{x}|\mathbf{z}) = \log_{\mathbf{x}_0}(\mathbf{x}_1)  \quad {\rm with}\quad \mathbf{z} = (\mathbf{x}_0, \mathbf{x}_1), \ \mathbf{x}_0\sim p, \ \mathbf{x}_1\sim \mu,
\end{align}
where the explicit form of the logarithmic map depends on the shape of the simulation box (see Appendix~\ref{app:logmap}).

\section{Implementation}
RFM as discussed above provides a general framework for modeling condensed-phase systems, while leaving considerable freedom in the precise implementation. In the following section, we specify our particular choices for the vector field and the prior distribution.

\textbf{Symmetries and Prior Alignment. }
The equilibrium configurational distribution naturally inherits the symmetries of the potential energy, including invariance wrt. global translations, rotations, and particle permutations. 
Embedding those symmetries into generative flow models has been demonstrated to substantially improve data efficiency and sample quality~\citep{kohler_equivariant_2020, satorras_equivariant_2022, klein_equivariant_2023}. In principle, the presented framework allows for straightforward incorporation of such symmetries, by combining a symmetry-invariant prior, such as the uniform distribution on the torus, with an equivariant vector field, as done in \citet{grenioux_riemannian_2025}. 
However, for the specific case of crystalline solids, we exploit the underlying lattice structure to adopt a distinct, physics-informed strategy: A 3$N$-dimensional wrapped Gaussian centered at equilibrium lattice sites~\citep{wirnsberger_normalizing_2022} (see Appendix~\ref{app:ec}) known as Einstein crystal~\citep{frenkel_new_1984}.

As the chosen lattice prior is not translation invariant, we restrict the system to a mean-free subspace, by fixing the center of mass of both the prior and target data.
The learned dynamics are constrained to preserve the center of mass by subtracting the mean particle velocity from both the learned vector field and the conditional target field.
The continuous $\mathrm{SO}(3)$ symmetry of Euclidean systems is reduced to a discrete subgroup for periodic boundary conditions.
Our lattice-based prior implicitly enforces a global alignment with the simulation box axes and a trivial assignment between prior and target samples. While we perform mini-batch optimal transport reordering~\citep{tong_improving_2023, pooladian_multisample_2023}, we omit further searches over discrete rotations and only adapt the transport cost calculation to torus distances.

\textbf{Vector Field Parameterization.}
To scale effectively to larger systems, we adopt a local modeling approach for the vector field, which ensures that  the learned dynamics are transferable across different system sizes~\citep{schebek_scalable_2025}.
We leverage the architectural flexibility of CNFs to adapt a powerful, size-transferable representation of atomic environments commonly used in machine learning interatomic potential architectures --- specifically the \textit{Equivariant Transformer} (ET)~\citep{pelaez_torchmdnet_2024} as proposed in \citet{hassan_etflow_2024} --- and consequently term our method RFM-ET in the following. 
Details on adaptations and hyperparameters are in Appendix~\ref{app:implementation}.

\textbf{Density Estimation.} 
%
\begin{figure}[]
    \centering
    \begin{subfigure}[t]{0.3\textwidth}
        \vspace{0pt}
        \centering
        \includegraphics[width=\linewidth]{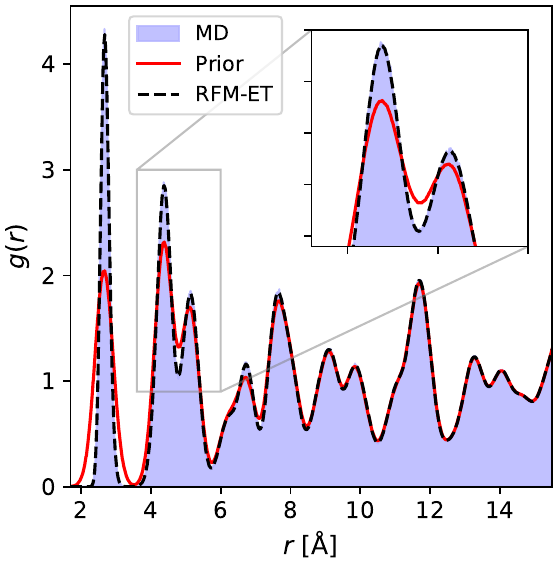}
    \end{subfigure}
    \hfill 
        \begin{subfigure}[t]{0.3\textwidth}
        \vspace{0pt}
        \centering
        \includegraphics[width=\linewidth]{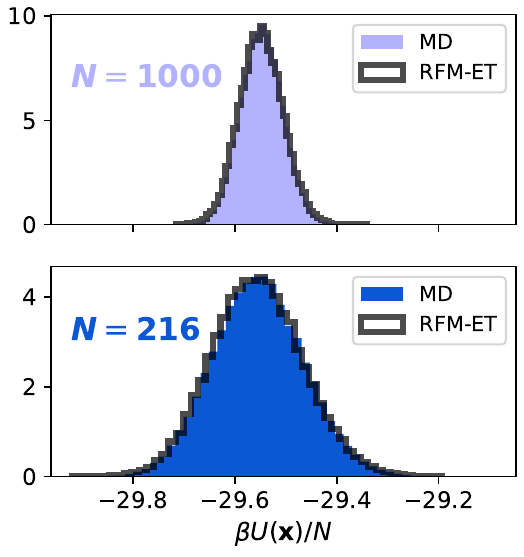}
    \end{subfigure}
    \hfill 
    \begin{subfigure}[t]{0.38\textwidth}
        \vspace{6pt}
        \centering
        \includegraphics[width=\linewidth]{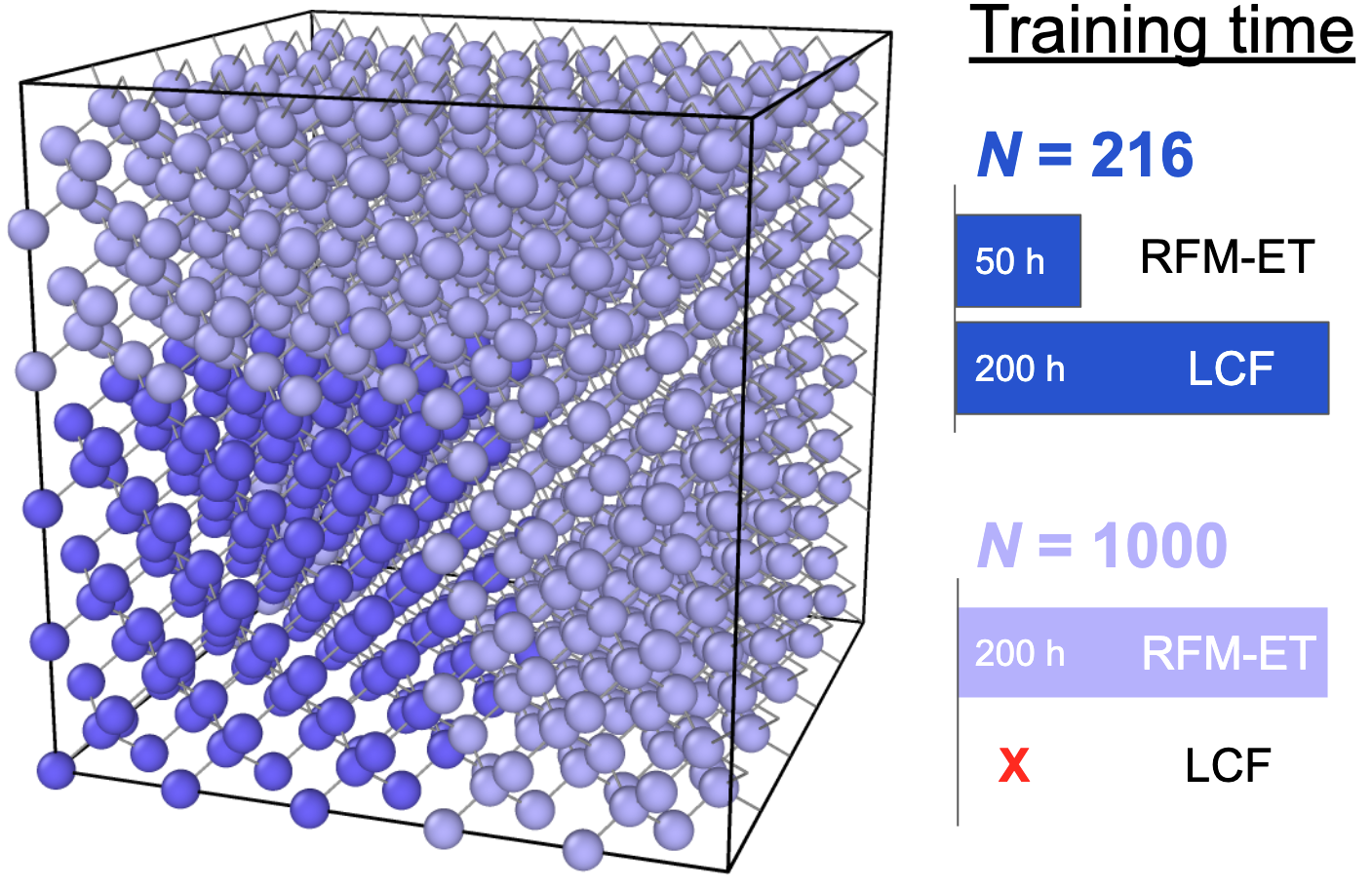}
    \end{subfigure}
    \caption{
    Left: Radial distribution function of the cubic ice system with 1000  particles as obtained from MD and RFM-ET.
    Center:  Energy histograms of the cubic ice system with 1000 (top) and 216 (bottom) particles as obtained from MD and RFM-ET.
    Right: The 216 particles (blue) and 1000 particles (light blue)  cubic ice systems and their training times with our method and local coupling flows (LCF)~(\citet{schebek_scalable_2025}).}
    \label{fig:rdf_ehist_system}
\end{figure}
As the chosen vector field parametrization 
naturally incorporates the toroidal topology and its outputs are directly in the Euclidean tangent space $\mathcal{T}_\mathbf{x} \mathcal{M} = \mathbb{R}^{3N}$, the Riemannian divergence simplifies to the standard Euclidean trace of the Jacobian, and we can use off-the-shelf ODE solvers, simply by projecting the final state back onto the torus.
Due to the GNN's local cutoff, the computational cost of evaluating the CNF vector field scales linearly with system size, $\mathcal{O}(N)$. However, calculating the exact divergence requires computing the trace of the full Jacobian. Because the input dimension $D$ is proportional to $N$, obtaining this trace requires $\mathcal{O}(N)$ evaluations of the vector field. This results in an overall $\mathcal{O}(N^2)$ complexity for both computation and memory, making the exact calculation infeasible for large systems.

Hutchinson's trace estimator \citep{Hutchinson1989ASE} offers an $\mathcal{O}(N)$ alternative but is often considered unsuitable for Boltzmann sampling applications due to noise and statistical bias.
Although Hutchinson's method provides an unbiased estimate of the vector field's divergence and thus the log-density change $\Delta \log p$, thermodynamic reweighting relies on the exponentiated log-density change.
Due to Jensen's inequality~\citep{jensen_fonctions_1906}, the nonlinearity of the exponential function transforms the estimator's variance into a systematic bias in the importance weights. This bias propagates to all downstream equilibrium estimates, such as free energy differences.
To address this, we propose a bias correction based on the second-order cumulant expansion~\citep{zwanzig_hightemperature_1954}.
By treating the stochastic error from the trace estimator as a fluctuating work term, we can expand the expectation of the likelihood estimate. Because the total accumulated noise arises from the sum of independent stochastic evaluations over multiple integration steps, its distribution rapidly converges to a Gaussian via the central limit theorem. Consequently, cumulants higher than the second order vanish, justifying the truncation of the series.
For each sample, we estimate the variance of the stochastic log-density change, $\hat{\sigma}^2$, and apply a correction term directly to the log-weights:
\begin{equation}
\widehat{\log w_{\text{corr}}}(\mathbf{x}_0) \coloneq \widehat{\log w}(\mathbf{x}_0) - \frac{1}{2}\hat{\sigma}^2(\mathbf{x}_0).
\end{equation}
This effectively removes the bias introduced by the estimator noise, allowing for accurate thermodynamic estimates.
To efficiently evaluate the correction term, we augment the ODE state with the empirical variance of the trace estimator, integrating it alongside the positions and log density.
A more detailed discussion is provided in Appendix~\ref{app:bias-correction} and Appendix~\ref{app:ode}.

\section{Experimental Results}
\begin{figure}[]
\begin{subfigure}[b]{0.66\textwidth}
    \centering
    \includegraphics[width=\linewidth]{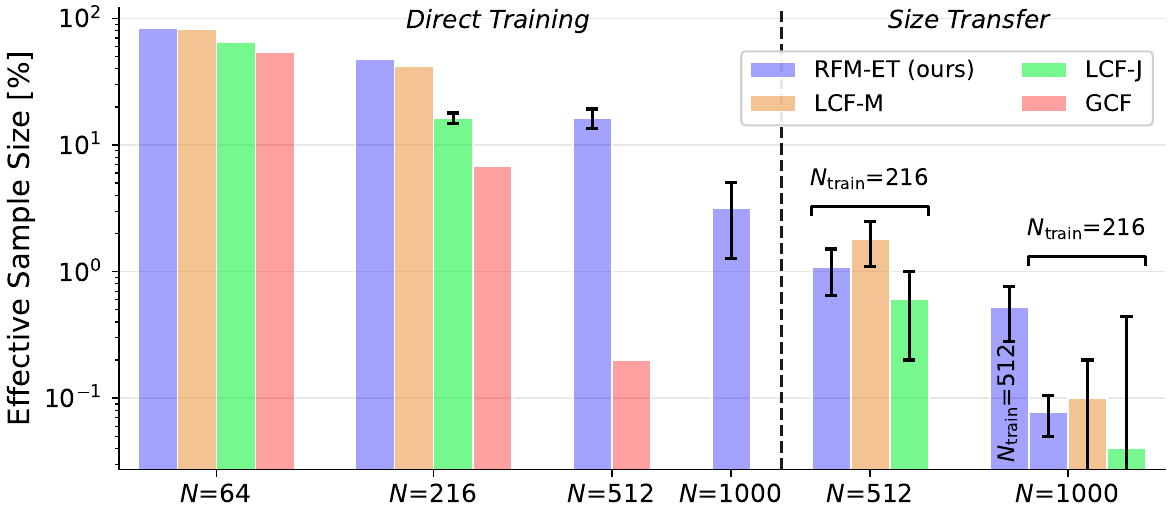}
\end{subfigure}
\hspace{0.001\textwidth}
\hspace{0.001\textwidth}
\begin{subfigure}[b]{0.31\textwidth}
    \centering
    \includegraphics[width=\linewidth]{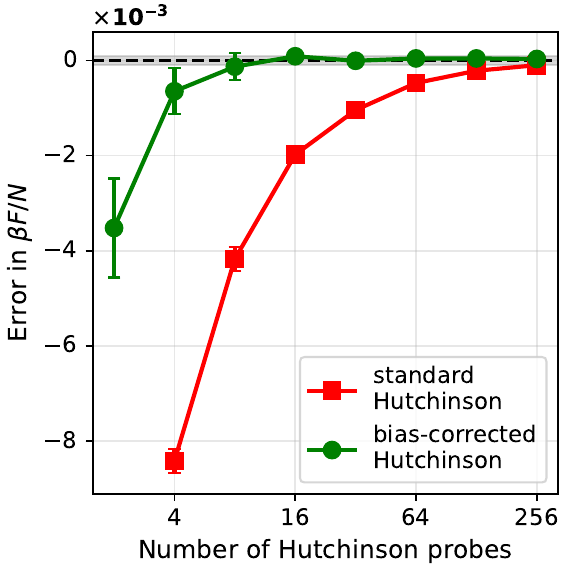}
\end{subfigure}

    \caption{
        Left: Effective sample size computed for the models described in the text, including results for models evaluated on the same system size as used during training, as well as for models transferred to larger system sizes than those seen during training. Error bars are shown where visible.    Right: The deviation to the reference free energy of a 512 particles mW system in the cubic ice phase, computed with a varying number of Hutchinson probes, with and without bias correction. The gray shaded area indicates a tolerance window of $10^{-4}$. Errors for our method in all plots were estimated over $10$k samples obtained from three models trained with different seeds.
    }
    \label{fig:ess_and_bias}
\end{figure}

We evaluate the performance of our method regarding both the quality of the generated samples and its accuracy as a free energy estimator. As a test system, we consider the common benchmark system cubic ice in the monoatomic water (mW)~\citep{molinero_water_2009, wirnsberger_normalizing_2022} parametrization. Figure~\ref{fig:rdf_ehist_system} (left and center) shows radial distribution functions and histograms of the potential energy as obtained from MD and RFM-ET for system sizes of 216 and 1000 particles. Importantly, even without any reweighting, excellent agreement is achieved for both observables, demonstrating that RFM-ET can accurately approximate the target Boltzmann distribution. 

The strong performance of RFM-ET is further reflected in the high ESS values shown in the left panel of Fig.~\ref{fig:ess_and_bias}. 
We first evaluate the performance of our model in the non-transferable setting, with all models assessed on the system size they were trained on. Where available, we compare to literature results obtained using a global transformer-based coupling flow (GCF)~\citep{wirnsberger_normalizing_2022} and a local augmented coupling flow (LCF)~\citep{schebek_scalable_2025}; for the LCF, estimates based on joint (LCF-J) and marginal (LCF-M) densities are included.
RFM-ET consistently matches or outperforms the others, demonstrating its high sampling quality.
Notably, training on a system size of $N=512$ yields a sampling efficiency two orders of magnitude higher than that of GCF, and, unlike the other models for which $N=1000$ is computationally infeasible, RFM-ET still achieves an ESS above three percent on this system size. Importantly, RFM-ET significantly reduces training times and enables training on large system sizes, such as $N=1000$, using the same computational budget that LCF requires to train on only $N=216$ (Fig.~\ref{fig:rdf_ehist_system}, right).

A crucial feature of our method is its transferability to larger system sizes, allowing models trained on small systems to be evaluated on larger systems without requiring additional data or training, similarly to the LCF architecture. As shown in the center of Fig.~\ref{fig:ess_and_bias}, RFM-ET matches the sampling efficiency of LCF-M when transferring models trained on $N=216$ to $N=512$ and $N=1000$, but only RFM-ET can be trained efficiently on $N=512$, which allows increasing the sampling efficiency for $N=1000$ by more than fivefold.

The right panel of Fig.~\ref{fig:ess_and_bias}  demonstrates the efficiency of the proposed bias correction by showing the relationship between the number of Hutchinson probes used per integration step and the free energy estimate obtained via TFEP. While the variance of the free energy estimates decreases rapidly with increasing number of probes, systematic bias remains the dominant source of error when no correction is applied. Even with 256 probes, the free energy estimated using biased weights is underestimated. In contrast, with as few as 16 probes, the bias-corrected weights allow the free energy estimates to converge to an accuracy of around $10^{-4}$, enabling resolution of free energy differences on the order of $10^{-3}$ between hexagonal and cubic ice under the given settings~\citep{schebek_scalable_2025}.
With fewer probes, the variance of the log-density change is underestimated, resulting in an insufficient bias correction.



\section{Discussion}
We showed that Riemannian Flow matching provides a powerful tool for equilibrium sampling of condensed matter, exhibiting superior scaling during training compared to coupling flows, enabling training on system sizes that were previously infeasible. Our method improves sampling on directly trained systems and enhances performance when models are transferred to larger systems. While the results presented here focus on cubic ice, we recently showed that the methodology is also applicable to other crystal structures within the mW potential and to systems described by the Lennard-Jones potential~\citep{schebek_assessing_2025}.
As also noted in the same study, two primary challenges remain. First, inference cost is high due to the computationally expensive ODE integration and stochastic divergence estimation required for density evaluation, making it one to two orders of magnitude slower than sampling with coupling flows. Second, although flow-matching formulations are generally robust even with limited training data, the method still relies on high-quality  configurations of the target distribution,  preventing the amortization of training costs through generalization over external conditions~\citep{schebek_efficient_2024}. A promising direction is data-free training of continuous-time models, which is currently being actively explored~\citep{pmlr-v267-havens25a, blessing2026bridgematchingsamplerscalable}.
Future work could further explore the integration of consistency distillation~\citep{geng2025improvedmeanflowschallenges} to enable few-step sampling without sacrificing the benefits of the continuous-time formulation, as well as the extension to the $NPT$ ensemble and physically more realistic potentials.

\subsubsection*{Acknowledgments}
MS acknowledges financial support from Deutsche Forschungsgemeinschaft (DFG) through grant CRC 1114 “Scaling Cascades in Complex Systems”, Project Number 235221301, Project B08 “Multiscale Boltzmann Generators”. The Flatiron Institute is a division of the Simons Foundation. We thank Michael Plainer for fruitful discussions and for reviewing the manuscript.


\bibliography{references_manual_v2}
\bibliographystyle{iclr2026_conference}

\newpage
\appendix
\section{Appendix}
\subsection{Geodesic on the flat torus} \label{app:logmap}

Let $\mathbf{x}_0, \mathbf{x}_1 \in \mathbb{R}^3$ be two points inside a periodic simulation box. The \emph{shortest vector} connecting them under periodic boundary conditions corresponds to the geodesic, which can defined by the log map 
$\log_{\mathbf{x}_0}(\mathbf{x}_1) \in \mathbb{R}^3$
with corresponding distance
$d = |\log_{\mathbf{x}_0}(\mathbf{x}_1)| \in \mathbb{R}$~\citep{gallot1990riemannian}
The concrete form of this log map depends on the shape of the simulation box~\citep{FrenkelSmit2023}:

\textbf{General triclinic box}\quad Let $\mathbf{A} \in \mathbb{R}^{3\times 3}$ be the matrix of lattice vectors,
\begin{equation}
\mathbf{A} = [\mathbf{a}_1, \mathbf{a}_2, \mathbf{a}_3], \quad \mathbf{a}_i \in \mathbb{R}^3.
\end{equation}
The log map is then
\begin{align}
\mathbf{s} &= \mathbf{A}^{-1} (\mathbf{x}_1 - \mathbf{x}_0) \in \mathbb{R}^3, \\
\mathbf{s}_\mathrm{min} &= \mathbf{s} - \Bigl\lfloor \mathbf{s} + \frac{1}{2} \Bigr\rfloor \in \mathbb{R}^3, \\
\log_{\mathbf{x}_0}(\mathbf{x}_1) &= \mathbf{A} \, \mathbf{s}_\mathrm{min} \in \mathbb{R}^3.
\end{align}

Throughout, $\lfloor \cdot \rfloor$ denotes the floor function applied component-wise.

\textbf{Orthorhombic box} \quad Let $\mathbf{L} = (L_x, L_y, L_z) \in \mathbb{R}^3$ contain the box lengths along each axis. Then the log map reads
\begin{equation}
\log_{\mathbf{x}_0}(\mathbf{x}_1) = (\mathbf{x}_1 - \mathbf{x}_0) - \mathbf{L} \cdot \Biggl\lfloor \frac{\mathbf{x}_1 - \mathbf{x}_0}{\mathbf{L}} + \frac{1}{2} \Biggr\rfloor \in \mathbb{R}^3,    
\end{equation}

where division and multiplication  are understood component-wise.

\textbf{Cubic box}\quad For a cubic box of side $L \in \mathbb{R}$, the formula simplifies to
\begin{equation}
\log_{\mathbf{x}_0}(\mathbf{x}_1) = (\mathbf{x}_1 - \mathbf{x}_0) - L \, \Biggl\lfloor \frac{\mathbf{x}_1 - \mathbf{x}_0}{L} + \frac{1}{2} \Biggr\rfloor \in \mathbb{R}^3.
\end{equation}

\subsection{Systems}

\subsubsection{3$N$-dim Gaussian prior: Einstein-Crystal} \label{app:ec}
The \(3N\)-dimensional Gaussian prior employed in this work corresponds to the configurational distribution of the so-called \textit{Einstein crystal}, i.e., a system of \(N\) independent harmonic oscillators with spring constant \(\Lambda_E\), centered on equilibrium lattice positions \(\{\mathbf{X}^i\}_{i=1}^N\) and confined to a box of volume \(V\). The corresponding potential energy is given by
\begin{equation}
  U_{\mathrm{id}}(\mathbf{x}) = \Lambda_E \sum_{i=1}^N \left|\mathbf{x}^i - \mathbf{X}^i \right|^2 .
\end{equation}

The partition function for this system can be analytically calculated and thus also its absolute free energy. Assuming a  fixed center of mass, one obtains
\begin{equation}
  \frac{\beta F_0}{N} = \frac{1}{N} \ln\left(\frac{N \Lambda^3}{V}\right) + \frac{3}{2}\left(1 - \frac{1}{N}\right) \ln\left(\frac{\beta \Lambda_E \Lambda^2}{\pi}\right) - \frac{3}{2N} \ln N,
\end{equation}
where \(\Lambda\) is the de Broglie wavelength, which is commonly set to a characteristic length scale of the interaction potential in model systems. The second term accounts for the constraint of fixing the center of mass~\citep{frenkel_new_1984}.

By computing the free energy difference between the Einstein Crystal and any arbitrary system, one can obtain the absolute free energy of the target system, which allows ranking different phases of a given system by stability.
We follow \citet{wirnsberger_normalizing_2022} by setting the width of the Gaussians, which is related to $\Lambda_E$, similar to the width of the displacements of the atoms around their equilibrium positions, with the exact widths provided in Tab.~\ref{tab:training_params}. For us, this ensures that the vector field always acts on a similar environment for each time $t$. The widths for the different systems are collected in Tab.~\ref{tab:training_params}.

\subsubsection{Monoatomic-Water} \label{app:mw}
The monoatomic water is a special case of the Stillinger-Weber (SW) potential~\citep{stillinger_computer_1985}. It incorporates two-body terms ($\phi_2$) as well as three-body ($\phi_3$) interaction terms, where the latter promote tetrahedral coordination environments. The total potential energy is expressed as:
\begin{equation}
U_{\mathrm{SW}}(\mathbf{x}) = \sum_i \sum_{j > i} \phi_2(d_{ij}) + \lambda_3 \sum_i \sum_{j \neq i} \sum_{k > j} \phi_3(d_{ij}, d_{ik}, \theta_{ijk}),
\end{equation}
where the individual interaction terms are defined as:
\begin{align}
\phi_2(r) &= A \varepsilon \left[ B \left(\frac{\sigma}{r}\right)^4 - 1 \right] \exp\left( \frac{\sigma}{r - a \sigma} \right), \\
\phi_3(r, s, \theta) &= \lambda \varepsilon ( \cos \theta - \cos \theta_0 )^2 \cdot \exp\left( \frac{\gamma \sigma}{r - a \sigma} \right) \cdot \exp\left( \frac{\gamma \sigma}{s - a \sigma} \right).
\end{align}

Here, $d_{ij}$ represents the distance between particles $i$ and $j$, and $\theta_{ijk}$ is the bond angle formed by the triplet of atoms $i$, $j$, and $k$. Most parameters in the SW potential are held fixed, with only $\varepsilon$, $\sigma$, and $\lambda_3$ being adjustable. The parameter $\lambda_3$ controls the strength of the three-body interactions, while $\varepsilon$ and $\sigma$ define the characteristic energy and length scales, respectively. The specific parameter values employed in this work follow those established in \citet{molinero_water_2009}, effectively modeling a coarse-grained representation of water.

Simulations were performed with the same settings as in \citet{wirnsberger_normalizing_2022} and \citet{schebek_scalable_2025} in the $NVT$ ensemble using the openMM package~\citep{Eastman2023}. The mW ice was modeled at $T = 200$\,K at a density of $\rho = 1.004$\,g\,cm$^{-3}$.

\subsection{Second-Order Cumulant Expansion for Hutchinson’s Estimator} \label{app:bias-correction}
To scale to large systems, we circumvent the effective $\mathcal{O}(N^2)$ complexity of the exact divergence calculation by employing Hutchinson’s trace estimator. At each integration step $t$, we approximate the trace using $K$ stochastic noise vectors  $\{\boldsymbol{\epsilon}_k \}_{k=1}^K$ with Rademacher entries, i.e., $\epsilon_{k,i} \in \{-1, +1\}$ i.i.d. with equal probability 1/2:
\begin{equation}
    \text{Tr}(\nabla v_\theta) \approx \frac{1}{K} \sum_{k=1}^K \boldsymbol{\epsilon}_k^T \nabla v_\theta \boldsymbol{\epsilon}_k.
\end{equation}
We denote the stochastic estimate of the log-density change obtained via this estimator as $\widehat{\Delta \log p}$. This estimate is \textit{unbiased}, meaning $\mathbb{E}[\widehat{\Delta \log p}] = \Delta \log p$.
This results in unbiased \textit{log} importance weights:
\begin{equation}
    \widehat{\log w}(\mathbf{x}_0) = \beta_p U_p(\mathbf{x}_0) - \beta_\mu U_\mu(f_\theta(\mathbf{x}_0)) - \widehat{\Delta \log p}.
\end{equation}
However, the computation of thermodynamic observables, such as the free energy difference, depends on the \textit{exponentiated} log weights.
By applying Jensen’s inequality to the expectation over the Hutchinson probes $\mathbb{E}_{\boldsymbol{\epsilon} \sim p(\boldsymbol{\epsilon} )}$, we see that:
\begin{equation}
    \mathbb{E}_{\boldsymbol{\epsilon} \sim p(\boldsymbol{\epsilon} )}[\hat{w}(\mathbf{x}_0)] = C(\mathbf{x}_0) \cdot \mathbb{E}_{\boldsymbol{\epsilon} \sim p(\boldsymbol{\epsilon} )}[\exp(\widehat{\Delta \log p})] \ge C(\mathbf{x}_0) \cdot \exp(\mathbb{E}_{\boldsymbol{\epsilon} \sim p(\boldsymbol{\epsilon} )}[\widehat{\Delta \log p}]) = w(\mathbf{x}_0),
\end{equation}
where $C(\mathbf{x}_0)$ contains the deterministic potential energy terms. This inequality implies that for any given sample $\mathbf{x}_0$, the stochastic weight is systematically overestimated on average. 
This bias propagates to the free energy estimator, which involves a second expectation over the samples $\mathbf{x}_0$ drawn from the prior $p(\mathbf{x}_0)$.
The estimator for the exponential of the free energy difference behaves as follows:
\begin{equation}
    \mathbb{E}_{\mathbf{x}_0 \sim p(\mathbf{x}_0)} \left[ \mathbb{E}_{\boldsymbol{\epsilon} \sim p(\boldsymbol{\epsilon} )} [\hat{w}(\mathbf{x}_0)] \right] \ge \mathbb{E}_{\mathbf{x}_0 \sim p(\mathbf{x}_0)}[w(\mathbf{x}_0)] = e^{-\beta \Delta F}.
\end{equation}
Consequently, taking the negative logarithm reverses the inequality, showing that unbiased stochastic trace estimation leads to a systematic underestimation of the free energy difference:
\begin{equation}
    \beta \Delta \hat{F} \coloneq -\log \mathbb{E}_{\mathbf{x}_0 \sim p(\mathbf{x}_0), \boldsymbol{\epsilon} \sim p(\boldsymbol{\epsilon} )} [\hat{w}(\mathbf{x}_0)] \le -\log \mathbb{E}_{\mathbf{x}_0 \sim p(\mathbf{x}_0)}[w(\mathbf{x}_0)] = \beta \Delta F.
\end{equation}

To address this, we treat the stochastic error from the trace estimator as a fluctuating work term and apply the \textit{second-order cumulant expansion}, a method rooted in thermodynamic perturbation theory~\citep{zwanzig_hightemperature_1954}. 
The expected value of the likelihood estimate can be expressed via the cumulant generating function expansion:
\begin{align}
    \log \mathbb{E}[e^{\widehat{\Delta \log p}}] &= \sum_{n=1}^\infty \frac{\kappa_n}{n!} \\
     &= \mathbb{E}_{\boldsymbol{\epsilon} \sim p(\boldsymbol{\epsilon} )}[\widehat{\Delta \log p}] + \frac{1}{2} \bigg(\mathbb{E}_{\boldsymbol{\epsilon} \sim p(\boldsymbol{\epsilon} )}[(\widehat{\Delta \log p})^2] -\mathbb{E}_{\boldsymbol{\epsilon} \sim p(\boldsymbol{\epsilon} )}[\widehat{\Delta \log p}]^2 \bigg) + ...
\end{align}
We truncate this series after the second order. This approximation is justified by the Central Limit Theorem: because the total accumulated noise arises from the sum of independent stochastic evaluations over many integration steps, its distribution converges rapidly to a Gaussian, even if the individual noise terms are non-Gaussian.
Since all cumulants $\kappa_n$ with $n > 2$ vanish for a Gaussian distribution, we correct the bias in the log-weights using the accumulated variance $\hat{\sigma}^2$ of the trace estimator along the trajectory. Since each sample $\mathbf{x}_0$ experiences a unique realization of the stochastic noise, both the log-density estimate $\widehat{\Delta \log p}$ and the variance correction $\hat{\sigma}^2$ are sample-specific:
\begin{equation}
    \widehat{\log w_{\text{corr}}}(\mathbf{x}_0) \coloneq \widehat{\log w}(\mathbf{x}_0) - \frac{1}{2}\hat{\sigma}^2(\mathbf{x}_0).
\end{equation}
The total variance $\hat{\sigma}^2$ is estimated by accumulating the local variance of the $K$ probes during the integration, as detailed in Appendix~\ref{app:ode}.

\begin{wrapfigure}{r}{0.33\textwidth}
    \centering
    \vspace{-\baselineskip}
    \includegraphics[width=\linewidth]{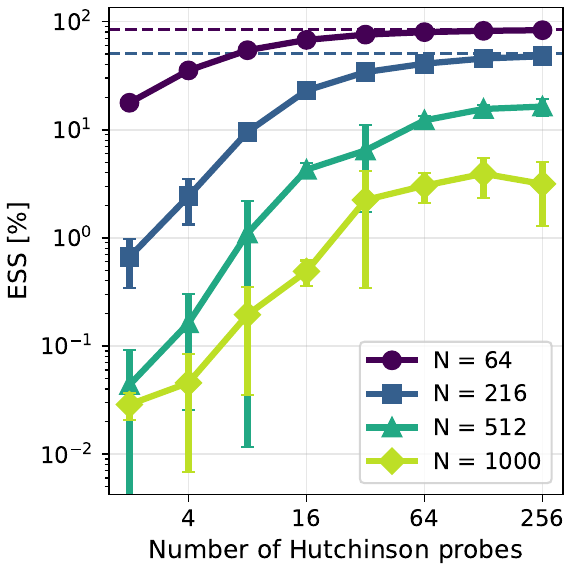}
    \caption{ESS against the number of Hutchinson probes used at each integration step. Dotted lines correspond to values obtained with exact divergence calculations.}
    \label{fig:ess_vs_k}
    \vspace{-\baselineskip}
\end{wrapfigure}

It is worth noting that for very few integration steps (e.g., $N < 5$), the Gaussian assumption may not hold. In such regimes, the noise follows the specific distribution of the Hutchinson estimator. When using Rademacher probes ($\mathbf{\epsilon} \in \{-1, 1\}$), the estimator possesses desirable sub-Gaussian concentration properties and minimizes the variance for specific matrix classes compared to Gaussian probes~\citep{avron_randomized_2011}. Importantly, the Rademacher distribution is symmetric, implying the skewness ($\kappa_3$) is zero. Therefore, if the Gaussian approximation fails due to a low step count, the leading error term in our expansion would be determined by the fourth cumulant $\kappa_4$ (kurtosis), rather than the third.

Finally, the stochasticity of the trace estimator directly impacts sampling efficiency. Since the importance weights inherit the noise from the $\Delta \log p$ estimate, the ESS inevitably degrades when fewer probes are used. Figure~\ref{fig:ess_vs_k} demonstrates that with a sufficient number of probes, the ESS converges to the baseline value obtained from exact divergence calculations. It is crucial to note, however, that the validity of our correction relies on an accurate estimate of the noise variance $\hat{\sigma}^2$. Using too few probes may yield an unreliable variance estimate, which can distort the correction term itself and result in residual bias.
We note that the ESS is negligibly affected by the bias-correction.

\subsection{ODE Integration} \label{app:ode}
We perform the integration of the CNF dynamics using a fixed-step 4th-order Runge-Kutta (RK4) solver. Adaptive step-size solvers are ill-suited for this application, as the stochastic nature of the Hutchinson trace estimator introduces high-frequency noise into the derivative of the log-density. Adaptive solvers typically interpret this statistical variance as local truncation error, leading to excessive and inefficient reduction of the step size. One could, however, use them for the positions only, but we found 10 RK4 steps to be sufficient in eliminating numerical integration errors, both in the positions and exact divergence calculations.

To estimate the accumulated variance $\sigma^2$ required for the cumulant correction without the computational cost of repeated trajectory realizations, we further augment the ODE state. We compute the empirical variance of the trace estimator, $\widehat{\text{Var}}[\widehat{\text{Tr}}]$, efficiently at each evaluation using $K$ probes and integrate it alongside the density:
\begin{equation}
    \frac{d}{dt} \begin{bmatrix} \mathbf{x} \\ \Delta \log p \\ \mathcal{V} \end{bmatrix} = \begin{bmatrix} v_\theta(\mathbf{x}, t) \\ -\widehat{\text{Tr}}(\nabla v_\theta) \\ \widehat{\text{Var}}[\widehat{\text{Tr}}(\nabla v_\theta)] \end{bmatrix}.
\end{equation}
By reusing the vector-Jacobian products computed for the trace, the overhead of computing the variance derivative is negligible.

However, the raw accumulated variance output $\mathcal{V}(T)$ must be corrected to account for the noise reduction inherent in the RK4 scheme. An RK4 step updates the log-density using a weighted average of derivatives $k_i$ evaluated at four stages:
\begin{equation}
    \log p_{t+h} = \log p_t + \frac{h}{6} (k_1 + 2k_2 + 2k_3 + k_4),
\end{equation}
where $h$ is the step size. Since the Hutchinson noise vectors are resampled at every stage, the noise terms in $k_i$ are independent random variables with local variance $\sigma_t^2$. The variance of the update step is therefore the sum of the variances of the weighted terms:
\begin{align}
    \text{Var}(\Delta \log p_{\text{step}}) &= \left(\frac{h}{6}\right)^2 \sum_{i=1}^4 a_i^2 \text{Var}(k_i) \\
    &\approx \frac{h^2}{36} (1^2 + 2^2 + 2^2 + 1^2) \sigma_t^2 \\
    &= \frac{10}{36} h^2 \sigma_t^2 = \frac{5}{18} h^2 \sigma_t^2.
\end{align}
The augmented ODE integrator accumulates the scalar rate, returning $\mathcal{V}(T) \approx \sum_{steps} \sigma_t^2 h$. To obtain the true variance of the path integral, we apply the correction:
\begin{equation}
    \sigma^2 = \mathcal{V}(T) \cdot h \cdot \frac{5}{18}.
\end{equation}

\subsection{Implementation details} \label{app:implementation}
We used the equivariant transformer from \citet{pelaez_torchmdnet_2024} and followed \citet{hassan_etflow_2024} by replacing the scalar output head with the equivariant output head from SchNet~\citep{schutt_schnet_2018} and using layer norms. Additionally, we added a bigger time dimension, removed the molecular features such as the bond specification and replaced the SMILES labeling by a sinusoidal encoding, based on the equilibrium position of an atom in its unit cell, as proposed in \citet{schebek_scalable_2025}. The hyperparameteres are collected in Tab.~\ref{tab:vector_field_params}.
For optimization the schedule free version of the AdamW optimizer~\citep{defazio_road_2024}, with the hyperparameters collected in Tab.~\ref{tab:optimizer_params}.

The work was implemented in \texttt{python}, using \texttt{pytorch}~\citep{paszke_pytorch_2019}, \texttt{lightning}~\citep{falcon_2019_pytorch}, \texttt{pytorch-geometric}~\citep{fey_fast_2019}, and \texttt{POT}~\citep{flamary_pot_2021}. To avoid waiting for mini-batch optimal transport reordering during training, it was reordered on the CPU simultaneously with training on the GPU.
The code is available under \url{https://github.com/emil-ho/bg-cnf-fm-condensed-matter}.

\begin{table}[htbp]
    \centering
    
    \caption{System-dependent Hyperparameters}
    \label{tab:training_params}
    \vspace{2mm}
    \begin{tabular}{@{} lcccc @{}}
        \toprule
        \textbf{Parameter} & \textbf{$\boldsymbol{N = 64}$} & \textbf{$\boldsymbol{N = 216}$} & \textbf{$\boldsymbol{N = 512}$} & \textbf{$\boldsymbol{N = 1000}$} \\
        \midrule
        Batch size & 128 & 128  & 128 & 128 \\
        Mini-Batch OT size & 128 & 128 & 64 & 32 \\
        Train steps & 800 K & 1 M  & 1 M & 1 M \\
        Prior width [\AA] & 0.1631 & 0.174 & 0.179 & 0.183 \\
        \bottomrule
    \end{tabular}

    \vspace{1cm} 

    
    \begin{minipage}[t]{0.48\textwidth}
        \centering
        \caption{Optimizer Hyperparameters}
        \label{tab:optimizer_params}
        \vspace{2mm} 
        \begin{tabular}[t]{@{} ll @{}}
            \toprule
            \textbf{Parameter} & \textbf{Value} \\
            \midrule
            Gradient clipping & 0.05 \\
            $\beta_1$ & 0.9 \\
            $\beta_2$ & 0.999 \\
            $\epsilon$ & $10^{-8}$ \\
            Learning rate & 0.0003 \\
            Warmup steps & 500 \\
            Weight decay & 0 \\
            \bottomrule
        \end{tabular}
    \end{minipage}
    \hfill 
    \begin{minipage}[t]{0.48\textwidth}
        \centering
        \caption{Vector Field Architecture}
        \label{tab:vector_field_params}
        \vspace{2mm} 
        \begin{tabular}[t]{@{} ll @{}}
            \toprule
            \textbf{Parameter} & \textbf{Value} \\
            \midrule
            Trainable params & 130 K \\
            Hidden channels & 24 \\
            Attention heads & 3 \\
            Layers & 12 \\
            Unit cell embedding & 1 \\
            Time embedding & 10 \\
            RBF features & 16 \\
            Activation & SiLU \\
            \bottomrule
        \end{tabular}
    \end{minipage}
    
\end{table}

\end{document}